# Addressing Gaps in Space Weather Operations and Understanding with Small Satellites

O. P. Verkhoglyadova[1], C.D. Bussy-Virat[2], Amir Caspi[3], D.R. Jackson[4], V. Kalegaev[5], J. Klenzing[6], J. Nieves-Chinchilla[7] and A. Vourlidas[8]

[1]Jet Propulsion Laboratory, California Institute of Technology, Pasadena, CA, USA

[2]Department of Climate and Space Sciences and Engineering, University of Michigan, Ann Arbor, MI, USA

[3]Southwest Research Institute, Boulder, CO, USA

[4]Met Office, UK.

[5]Skobeltsyn Institute of Nuclear Physics, Moscow State University, Moscow, Russia

[6]ITM Physics Laboratory, Goddard Space Flight Center, Greenbelt, MD, USA

[7] Consultant-5G/Sat Ground Stations Facilities, Spain

[8]Johns Hopkins University Applied Physics Laboratory, Laurel, MD, USA

Corresponding author: Olga Verkhoglyadova (Olga.Verkhoglyadova@jpl.nasa.gov )

**Key Points:**

- Enhancing space weather operations and understanding with small satellites is discussed
- Key observables and small satellite strategies are recommended

**Abstract**

Gaps in space weather observations that can be addressed with small satellites are identified. Potential improvements in solar inputs to space weather models, space radiation control, estimations of energy budget of the upper Earth's atmosphere, and satellite drag modeling are briefly discussed. Key observables, instruments and observation strategies by small satellites are recommended. Tracking optimization for small satellites is proposed.

1 **Introduction**

The updated National Space Weather Strategy and Action Plan formulates its Objective II as the task to develop and disseminate accurate and timely space weather characterization and forecasts, including regional and global characterization of space weather conditions (NSTC, 2019). Meeting the Objective requires space weather monitoring from ground and space (Knipp & Gannon, 2019). Due to continuing improvements in small satellite (SmallSat) technologies, SmallSat missions are gaining more interest as solutions to address long-standing scientific problems and operational needs (Moretto & Robinson, 2008; NASEM, 2016).

The World Meteorological Organization (WMO) produces requirements[1] for observations of physical variables in support of all WMO Programs, including space weather. The requirements are rolling and are regularly reviewed by the WMO Inter-Program Team on Space Weather Information, Systems and Services (IPT-SWeISS), whose members are experts typically representing national operational space weather centers. The requirements emphasize near-real time space weather operations. IPT-SWeISS assessments indicate that the WMO requirements are often poorly met by the existing observation network and gaps could be very effectively filled by observations from a SmallSat constellation.

The WMO requirements are a means by which improvements to space weather observations can be advocated, which requires good communication between forecasters, instrument developers, and researchers. However, much more work needs to be done to publicize the WMO requirements list, especially in the SmallSat community. By using the requirements as a focus for SmallSat design, we can work together more effectively to fill the gaps in the observational network, and to enable SmallSat observations to be increasingly useful for research and operational applications.

The 1st International Workshop on SmallSats for Space Weather Research and Forecasting (SSWRF), held in Washington, DC on 2017 August 1–4, brought together experts in heliophysics, space physics, space weather operations, and related fields to help identify how SmallSats could fill current gaps in space weather understanding and forecasting. Those findings are discussed here.

2 **Current Gaps and Recommendations**

SmallSats have a potential to enable cutting-edge heliophysics science (NASEM, 2016). A number of successful missions have already demonstrated such capability (Spence et al., 2020), and many future missions have been funded or proposed, including a number with direct relevance to space weather science and operations (e.g., Caspi et al., 2020; Klenzing et al., 2019). Some of the key advantages of using SmallSats for progressing heliophysics science are their high-heritage technology, rapid replaceability, low cost, and constellation opportunities for distributed measurements. These advantages can be utilized to fill the important data gaps listed below. Table 1 lists several key observables that can be achieved with instruments launched on SmallSats, and additional advances in critical enabling technologies will further enhance SmallSat capabilities for space weather operations (Klumpar et al., 2020). Justification and additional considerations for the observables are provided below.

---

[1] http://www.wmo-sat.info/oscar/applicationareas/view/25

## 2.1 Solar inputs of space weather models

One of the ongoing concerns for space weather operations is the reliance of forecasting models on data inputs from research-type instrumentation. The inputs are of a critical nature, such as the key parameters of solar flares (e.g., spectral irradiance) and CMEs (e.g., occurrence, initial speed, direction), which constitute the main drivers of space weather, and solar photospheric magnetic fields that are essential for establishing the background state of the heliosphere.

Solar telescope designs are mature with well understood measurement requirements. Many telescopes can be readily miniaturized without compromising their space weather value; some already have been. For example, MiniCOR (Korendyke et al., 2015), a 6U Cubesat-compatible version of the STEREO/SECCHI COR2 coronagraph, fulfills NOAA's operational requirements for a coronagraph, except for the extended lifetime requirement for operational payloads. The recently-selected PUNCH Small Explorer will also include a miniaturized coronagraph and three wide-field imagers, all on SmallSat platforms (see Caspi et al., 2020). Soft X-ray spectrometers such as those flown on MinXSS (Mason et al., 2016; Woods et al., 2017), a 3U CubeSat, can provide detailed spectral irradiance diagnostics of flares and active regions for better modeling of Earth ionospheric reactions to solar soft X-ray forcing (see also Sec. 2.3).

The limited lifetime and/or robustness of SmallSats can be addressed by frequent unit deployment and replacement via rideshare, and/or multi-unit deployment strategies. Small satellites can enable particularly novel inputs from EUV (information on the pre-eruptive structure and initial eruption stages) and soft and hard X-ray imagers (plasma composition and particle acceleration) from off-Sun-Earth line viewpoints (e.g., from $L_5$; Vourlidas, 2015). Although SmallSat-sized EUV and soft/hard X-ray imagers and imaging spectrometers, and Doppler magnetographs, have been discussed in the scientific community, no such instruments have yet been funded or flown. The development of such miniaturized designs would greatly benefit our understanding of space weather modeling inputs, with the ultimate goal of acquiring a standardized set of small satellite space weather sensors for rideshare deployment in Sun-synchronous polar orbits or interplanetary orbits (to $L_4$, $L_5$, or elsewhere).

## 2.2 Space radiation operational control

Space radiation poses major hazards for satellites including spacecraft charging (surface and internal), radiation dose, and single-event effects (Schrijver et al., 2015). Continuous monitoring of energetic particle fluxes in geospace from solar energetic particles (SEPs), galactic cosmic rays (GCRs), and radiation belt (RB) particles, as well as auroral low-energy particles, is crucial to mitigate space weather radiation risks in accordance with WMO directives. Increases of SEP or trapped energetic electron fluxes by 2–4 orders of magnitude can occur after solar flares or after passage of high-speed streams, respectively. Direct flux measurements are the most effective way to mitigate space weather risks. In-situ SEP event detections by existing space weather services are now performed by spacecraft located in the solar wind (e.g., ACE) or at Geostationary Earth Orbit (GEO). Multi-satellite observations by small satellites in low Earth orbit (LEO) can significantly improve SEP event detection and give information on the size of the polar area accessed by SEPs.

The outer (electron) radiation belt (ORB) is the most dynamical part of Earth's radiation environment. Relativistic and sub-relativistic ORB electrons contribute to total ionizing dose and internal charging, leading to electrostatic discharge that can damage spacecraft or destroy instrumentation. Existing empirical models of Earth's radiation belts (Vette, 1991; Ginet et al., 2013) are unable to reproduce significant short-term variations of the energetic electron fluxes during magnetospheric disturbances and cannot be used in operational space weather services. Numerical models like VERB (Shprits et al., 2009) are not so fast for real-time services. Reproduction of fluxes over the whole ORB requires in situ monitoring of all magnetic L-shells. This can be done by polar LEO satellites (e.g., POES, Jason, SAC-D, Meteor M), but particle flux distributions are quite complex: one needs simultaneous spatially-distributed measurements. Multiple measurements performed by identical low-cost SmallSats at different longitudes and L-shells provides knowledge of locations of high radiation risk areas and electron flux values.

Simultaneous multi-point observations provide excellent opportunities for early detection of the both SEP and ORB radiation event onsets as well as strong auroral particle precipitation. Radiation events can continue from several hours to several days. Early detections of SEP and ORB events give more time to protect satellite electronics against environment factors. SmallSat constellations could thus improve data coverage in a cost-effective, yet comprehensive, way not available via existing, more conventional methods. Data assimilation RB models (see Bourdarie et al., 2007) depend strongly on data volume and SmallSat data would be useful for overcoming deficiencies in such modeling.

Key parameters needed for space radiation monitoring are listed in Table 1. Several detectors with different orientations can measure pitch-angle distributions and reconstruct the particle fluxes in the whole inner magnetosphere. This is the approach of the Moscow State University "Universat-SOCRAT" project to develop a system of SmallSats for operational monitoring of radiation conditions in near-Earth space (Panasyuk et al., 2017).

## 2.3 *Constraining the energy budget of the Earth's upper atmosphere*

The Earth's ionosphere and upper neutral atmosphere are driven from above (solar wind and magnetosphere) and from below (middle and lower atmosphere). First-principles models of the coupled ionosphere-thermosphere (IT) system are the foundation of many forecasting efforts and rely on accurate driver specifications (Mannucci et al., 2016). Since the IT system is highly sensitive to driving (e.g., Siscoe & Solomon, 2006) uncertainties in energy inputs and energy budget drive large errors in modeling of IT state (Deng et al., 2013; Verkhoglyadova et al., 2017) and potentially its forecasting (Verkhoglyadova et al., 2020). Mannucci et al. (2020) emphasized the need for continuously available low latency observations directly relevant to space weather. There is growing observational evidence that without considering energy transport at multi-scales, energy input into the ionosphere may be underestimated (Chaston et al., 2005; Huang et al., 2017; Miles et al., 2018; Zhu et al., 2018; Ozturk et al., 2020).

There is a need for dedicated mission focused on understanding magnetic and electric field variability and providing energy inputs at mesoscales that will improve accuracy of first-principles models and their predictive capability. A SmallSat (or Cubesat) constellation targeting the high-latitude ionosphere between 100 and 600 km altitude and employing in situ (see Table 1) or remote sensing measurements (e.g., Yee et al., 2017) would provide an integration of SmallSat-based information and quantitative improvement of the space weather modeling framework.

Moreover, there is a lack of understanding of the solar soft X-ray spectral irradiance that is both variable (Rodgers et al., 2006) and strongly impacts the Earth's ionosphere-thermosphere-mesosphere (Sojka et al., 2013, 2014). Moderate-resolution measurements (e.g., Caspi et al., 2015; Moore et al., 2018) at both short (flaring) and long (solar rotation) timescales are needed to understand this critical energetic input to the Earth's upper atmosphere. Several suggested key observables are listed in Table 1.

### 2.4 *Improvement of satellite drag models*

Another area of space weather that can benefit from SmallSats is satellite drag estimation. Changes to the exospheric temperature, the local composition of neutral species, and to their net motion modify the atmospheric density, hence the drag on satellites. Errors in the modeling of drag have important consequences on missions, particularly in terms of collision avoidance (Bussy-Virat et al., 2018) and mission lifetime. However, these parameters have hardly been measured since Atmospheric Explorer and Dynamics Explorer. Global measurements of the neutrals and winds are missing, which affect the calibration of the models and the fidelity of predictions (e.g., Vallado & Finkleman, 2014). A WMO analysis of the existing observations network rated it in general to be poor for operational use.

Semi-empirical thermosphere models, such as the Drag Temperature Model (Bruinsma, 2015), are already used widely for operational calculations of satellite drag. While the results are reasonably good, there are questions regarding the consistency of the accelerometer observations on which they are largely based (e.g., March et al., 2019). Measurements of total neutral pressure, temperature, mass composition, and winds would therefore be useful in constructing the next generation of satellite drag models. These new data can also be added to data assimilation systems (e.g. Murray et al., 2015; Elvidge & Angling, 2019) which are being developed to run in conjunction with physics-based models and which potentially can supply better thermospheric analyses than semi-empirical models. The development of miniaturized instruments – such as neutral pressure sensors (Bishop et al., 2019), neutral mass spectrometers (Rodriguez et al., 2016), neutral wind meters (Earle et al., 2018), and Fabry-Perot Interferometers (Harlander & Englert, 2020) – can provide multiple opportunities to capture new information about the neutral atmosphere (see Table 1). Short-term missions to low altitudes (250–500 km) in the high-drag regime become more feasible as the relative cost of these platforms drops. Additionally, constellations of measurements can help capture the complex spatio-temporal reaction of the thermosphere to magnetic storm inputs, ranging from hundreds of km up to global scales. See Caspi et al. (2020) for a more detailed discussion.

Constellations of SmallSats equipped with GPS receivers can also improve the estimation of atmospheric density. By applying filtering techniques (Chen & Sang, 2016) to the ephemerides of each member of the constellation, small scale features and short temporal variations in the density can now be detected. For example, the CYclone Global Navigation Satellite System (CYGNSS) mission (Ruf et al., 2013), a constellation of eight SmallSats whose primary goal is to measure winds in cyclones, is used to estimate the density at ~500 km. In addition, high-drag maneuvers are carried out to control the trajectories of the CYGNSS satellites. These maneuvers imply different magnitudes of the drag force, thus the differing effects on the satellite trajectories can be used to improve the accuracy of the density estimation in the filtering algorithms.

## 2.5 *The role of ground support in tracking optimization for SmallSats*

Since the beginning of the CubeSat era around the year 2005, the platform capabilities and the mission coverage are the two main critical points which need to be addressed in the system communication design for LEO missions. In addition, spectrum congestion areas around Ground Station (GS) facilities should be considered for tracking optimization of a successful mission.

Increasing telemetry volume and decreasing data latency can enhance the use of SmallSats in LEO for space weather applications. Some improvements can be achieved through optimized site selection when deploying or choosing ground stations (GSs). For any single GS, the antenna elevation mask (i.e., view to the sky), customized for a mission's radio frequency band, is the most relevant and impactful parameter. Recently-developed 3D tools and techniques can accurately determine the elevation mask based on surrounding view obstruction and RF interference sources. This enables determination of optimal antenna placement, taking into account mission performance requirements and facility constraints, and can increase satellite visibility by up to 50% in built-up, urban environments (Nieves-Chinchilla et al., 2017, 2018). Utilizing a network of distributed GSs provides significant additional benefit, increasing available downlink by 5–10 times when the longitudinal distribution of the stations is optimized to minimize overlap and maximize visibility. For example, SatNet can help to construct the necessary telecom infrastructure to enable the sharing of radio amateur GSs between CubeSat operators (Tubío-Pardavila et al., 2016). Alternatively, some recent SmallSats have made use of real-time satellite-to-satellite communications (e.g., GlobalStar, Iridium) to enable real-time downlink separately from GS visibility, albeit at relatively low data rates. Klumpar et al. (2020) discuss additional technological advances in communications and data analytics that would improve SmallSat capabilities for space weather operational capacity, while Nieves-Chinchilla et al. (2020) discuss relevant issues that must be considered for international coordination, e.g., for frequency licensing and/or deployment and use of widespread GS networks.

Advances in CubeSat technology, design of constellation formations such as the LAICE CubeSat mission (Westerhoff et al., 2015) and the QB50 project (Gill et al., 2013), and utilizing a network of distributed GSs such as the SatNet project (Tubío-Pardavila et al., 2016) have positioned CubeSats as an alternative for space weather exploration. Feasibility of both real-time and non-real-time telemetry technology for SmallSat platforms needs to be explored. Low-latency or constant downlinking is critical for time-sensitive observations, e.g., alerts for solar flares, CMEs, and SEPs, while less critical, but more detailed, information can be downlinked in a traditional manner with higher latency. The need of data volume and the need to support SmallSat missions, tracking and control operations, has given rise to a major relevancy on GS locations and design in urban environments. The tracking optimization could maximize the satellite access times and therefore enhance space weather operations and understanding with SmallSats platforms.

## 3 Summary

We have identified several current gaps in space weather understanding and operational needs that can be addressed with SmallSats. We recommended key observables, instruments and observations strategies that can enhance space weather operations in several domains.

**Acknowledgments**

Portions of O.V. research were performed at the Jet Propulsion Laboratory, California Institute of Technology, under a contract with NASA. A.C. was partially supported by NASA grants NNX14AH54G, NNX15AQ68G, NNX17AI71G, and 80NSSC19K0287. A.V. was supported NRL grant N00173-16-1-G029. V.K. research on radiation monitoring was supported by RScF-16-17-00098 Grant. J.K. was supported by NASA H-TIDeS through ROSES NNH16ZDA001N. CBV was supported by NASA grant NNL13AQ00C. No new data were used in preparing this manuscript.

| Observables | Instruments | Small-Sat Strategy |
|---|---|---|
| Solar flare occurrence & intensity | Soft X-ray spectrometer (0.5–15 keV, ≥64 channels) | Small-to-medium constellation, 3–6U bus, sun-pointed, 3-axis stabilized; LEO; real-time satcomm (e.g., GlobalStar) for low-latency alerts & lightcurve downlinks |
| Soft X-ray irradiance & variability (forcing of Earth ITM) | FOV: ~1° [full disk + lower corona], spatially integrated, few-sec cadence | |
| CME occurrence | White light coronagraph (FOV: 3-20 $R_S$, spatial resolution: 30–60 arcsec; cadence: 15–30 min) | Adapt Mini-COR or CCOR design to a 6–12U bus. ESPA-compatible. Ruggedize for deep-space applications. Sun-Sync Polar or 1 AU drifting orbit. |
| CME speed at 20 $R_S$ | | |
| CME direction | | Same as above but deploy >30° from Sun-Earth line |
| Photospheric Magnetic Field | LOS magnetograph (FOV: full disk, spatial resolution: 2–4 arcsec) | 12U bus. Adapt Solar Orbiter/PHI data reduction FPGA code for onboard analysis. |
| Electron flux monitoring in the ORB and in auroral region | Particle detector (0.1–4 MeV) Particle detector (0.5–20 keV) | Altitude range: 600–2000 km, access to L > 10, time resolution <10 sec.; 3–4 energetic channels for each high-energy detector. 5–10 channels for low-energy detector. |
| SEP flux monitoring in polar cap | Particle detector (1–100 MeV) | |
| Magnetic field in the Earth's ionosphere | 3-component DC magnetometer, ULF frequency range from ~1 mHz to ~5 Hz | LEO high inclination orbit, altitude range of the ionospheric measurements:100–600 km |
| Electric field in the Earth's ionosphere | Electric field sensor, the same frequency range, magnitudes up to 200 mV/m. | |
| Total ion density, ion/electron temperatures in the Earth's ionosphere | Langmuir Probe, Retarding Potential Analyzer | |
| Upper atmosphere parameters (total neutral pressure, temperature, mass composition, winds, total electron content, flux of molecular oxygen) | Neutral pressure sensors, neutral mass spectrometers, neutral wind meters, flux probe | Short-term missions to low altitudes (250–500 km), high drag regime, ability to resolve from hundreds of km to global scales. |

**Table 1.** Key observables for space weather science and operations that can be provided with SmallSat missions.